# A dynamic black hole corona in an active galaxy through X-ray reverberation mapping

William N. Alston[1*], Andrew C. Fabian[1], Erin Kara[2], Michael L. Parker[3], Michal Dovciak[4], Ciro Pinto[5,6], Jiachen Jiang[1,7,8], Matthew J. Middleton[9], Giovanni Miniutti[10], Dominic J. Walton[1], Dan R. Wilkins[11], Douglas J. K. Buisson[9], Maria D. Caballero-Garcia[3], Edward M. Cackett[12], Barbara De Marco[13], Luigi C. Gallo[14], Anne M. Lohfink[15], Chris S. Reynolds[1], Phil Uttley[16], Andrew J. Young[17], Abderahmen Zogbhi[18]

[1]Institute of Astronomy, Madingley Rd, Cambridge, CB3 0HA.
[2]MIT Kavli Institute for Astrophysics and Space Research, Cambridge, MA 02139, USA
[3]European Space Agency (ESA), European Space Astronomy Centre (ESAC), E-28691 Villanueva de la Canada, Madrid, Spain
[4]Astronomical Institute of the Czech Academy of Sciences, Boční II 1401, CZ-14100 Prague, Czech Republic
[5]ESTEC/ESA, Keplerlaan 1, 2201AZ Noordwijk, The Netherlands
[6]INAF - IASF Palermo, Via U. La Malfa 153, I-90146 Palermo, Italy).
[7]Department of Astronomy, Tsinghua University, Beijing 100084, China
[8]Tsinghua Center for Astrophysics, Tsinghua University, Beijing 100084, China
[9]Department of Physics and Astronomy, University of Southampton, Highfield, Southampton, SO17 1BJ, UK
[10]Centro de Astrobiologia (CSIC–INTA), Dep. de Astrofisica; ESAC campus, E-28692 Villanueva de la Canada, Spain
[11]Kavli Institute for Particle Astrophysics and Cosmology, Stanford University, Stanford, CA 94305, USA
[12]Department of Physics & Astronomy, Wayne State University, 666 W. Hancock St, Detroit, MI 48201, USA
[13]Nicolaus Copernicus Astronomical Centre of the Polish Academy of Sciences, ul. Bartycka 18, 00-716 Warszawa
[14]Department of Astronomy and Physics, Saint Mary's University, 923 Robie Street, Halifax, NS B3H 3C3, Canada
[15]Montana State University, P.O. Box 173840, Bozeman, MT 59717-3840
[16]Astronomical Institute "Anton Pannekoek", University of Amsterdam, Science Park 904, 1098XH, Amsterdam, the Netherlands
[17]H. H. Wills Physics Laboratory, Tyndall Avenue, Bristol BS8 1TL
[18]University of Michigan, Department of Astronomy,1085 S. University, Ann Arbor, MI 48109

**X-ray reverberation echoes are assumed to be produced in the strongly distorted spacetime around accreting supermassive black holes. This signal allows us to spatially map the geometry of the inner accretion flow[1,2] - a region which cannot yet be spatially resolved by any telescope - and provides a direct measure of the black hole mass and spin. The reverberation timescale is set by the light travel path between the direct emission from a hot X-ray corona and the reprocessed emission from the inner edge of the accretion disc[3-6]. However, there is an inherent degeneracy in the reverberation signal between black hole mass, inner disc radius and height of the illuminating corona above the disc. Here, we use a long X-ray observation of the highly-variable active galaxy, IRAS 13224–3809, to track the reverberation signal as the system evolves on timescales of a day[7,8]. With the inclusion of all the relativistic effects, modelling reveals that the height of the X-ray corona increases with increasing luminosity, providing a dynamic view of the inner accretion region. This simultaneous modelling allows us to break the inherent degeneracies and obtain an independent timing-based estimate for the mass and spin of the black hole. The uncertainty on black hole mass is comparable to the leading optical reverberation method[9], making X-ray reverberation a powerful technique, particularly for sources with low optical variability[10].**

IRAS 13224–3809 is a nearby and bright active galactic nucleus (AGN). As a Narrow Line Seyfert 1 (NLS1) type AGN, it is characterised by a high rate of accretion[11,12] onto a relatively low mass, supermassive black hole ($M_{BH} \sim 10^6\ M_\odot$, where $M_\odot$ is the solar mass). In accordance with such extreme rates of accretion, highly ionised winds driven from the accretion flow at near relativistic speeds have been detected from this source[13,14]. IRAS 13224–3809 has been observed for 16 full orbits (~130 ks per orbit) with the European Space Agency's X-ray Multi-Mirror Mission (*XMM-Newton*[15]), totalling 2 Mega-seconds of observations. The source is one of the most variable X-ray objects in the sky, undergoing rapid and large amplitude variations on timescales of minutes[7,8]. It is unique in that the shape of the variability (visualised as power spectral density – PSD) is also time-dependent, varying on days timescales[7,8]. This short-timescale variability is observed in most AGN, which - along with gravitational microlensing observations of quasars[16] - indicates that the X-rays are emerging from a region within the central ~15 $R_g$ (where $R_g = GM_{BH}/c^2$ is the gravitational radius, $G$ is the gravitational constant and $c$ is the speed of light). At a distance of nearly a billion light years, such regions have an angular size on the sky too small to be resolved using present or planned instruments.

Observations of AGN tell us there must be a hot 'corona' of electrons close to the black hole, which constitutes a significant fraction of their total luminosity[11,12]. The origin and geometry of this corona is unknown, but it irradiates the inflowing matter, with gravitational light bending focusing most of the radiation on the inner edge of the accretion disc, before being reprocessed and directed toward the observer (see Figure 1). Such reflection produces a number of X-ray emission lines, the strongest of which is associated with iron, as well as intense X-ray emission lines below ~1 keV. The distortion of these lines encodes important information about the geometry of the inner accretion flow and the black hole via a host of relativistic effects[17]. Modelling such lines in IRAS 13224–3809 indicates that the black hole is rapidly spinning[18,19], having a dimensionless spin parameter, $a$ $(= Jc/GM_{BH})$ close to the maximal permitted value of unity. Spin also sets the position of the innermost stable circular orbit $R_{ISCO}$, within which, matter plunges into the black hole.

The physical separation between the corona and the accretion disc means that a distant observer will detect a time lag between the intrinsic emission (which dominates in the 1.2-5.0 keV band) and the soft reflection features (typically in the 0.3-1.0 keV band)[2,20,21]. As with the reflection spectrum, the time lags are also affected by the geometry of the inner accretion flow and associated relativistic effects. However, unlike the energy spectrum which provides size-scales in units of $R_g$, the lags provide the same information but in absolute physical units. Modelling these time lags is therefore a powerful and independent technique for understanding the details of the accretion flow and properties of the central black hole. To date, modelling time-lags in individual observations have yielded only a static picture[3–6,22] and degeneracies remain. For instance, in an individual observation there is some degeneracy between mass $M_{BH}$ and source height, $h$, because the characteristic time-lag scales as $\tau \sim hM_{BH}$: the same lag would be produced by a combination of low height and high mass, or low mass and large height.

Four earlier *XMM-Newton* observations of IRAS 13224–3809 hinted at an epoch-dependence to the time lags, likely due to physical changes in the geometry of the inner accretion flow[23]. The lag spectra from 12 new XMM-Newton observations also show large changes (see Extended Data Figure 1), with the frequency of the largest amplitude negative lag changing by as much as an order of magnitude in frequency. The strong and short timescale variability in IRAS 13224–3809 means we can model the time lags as they rapidly evolve, helping to break these degeneracies, whilst the number of observations allows us to create a movie for how the corona evolves.

We model the Fourier lag-spectra using a transfer function which can be seen as the response of the disc to a flash of X-rays coming from the corona, encoding all the effects of geometry, relativity and light bending[24] - see Figure 1. The disc-reflection scenario[17,18,19,20,22,25] assumes the corona to be a static, point-like X-ray source located at height $h$ on the rotation axis above the black hole, and which isotropically emits a power-law of flux with energy (flux $\propto E^{-\Gamma}$). This illuminates a geometrically thin, optically thick and ionised accretion disc which extends down to some radius $R_{in}$ (and which we fix to have an outer radius at $400R_g$), and which the observer sees at some inclination $i$ (where $i = 0°$ is a disc seen face-on). We fix the power-law index $\Gamma$ and $2 - 10$ keV luminosity, $L_{2-10\ keV}$, from fits to the time-averaged spectra for each observation, and which account for the presence of an ultrafast wind[14] (see Methods). From the time-averaged spectral fits, we are able to obtain the relative weighting of the reflection to the intrinsic power-law in each energy band; this sets the amount of dilution of the reflected signal in a given band[2] which affects the magnitude of the reverberation lag. In addition to dilution, a lag associated with the inward propagation of matter dominates at low frequencies ($< 5 \times 10^{-4}$ Hz) in NLS1 galaxies, which is included as an additional component in our modelling of the lag spectra (see Methods).

In the transfer function model fitting, $R_{in}$, $M_{BH}$, spin and inclination are free parameters and tied across all lag spectra. Two best-fitting model examples are shown in Figure 1, with the remainder shown in Extended Data Figure 1. We test for dynamical changes in the height of the corona by assessing the relationship between source height $h$ and the observed coronal luminosity $L_{2-10\ keV}$ (see Figure 2 and Extended Data Figure 2 and 3). A strong linear trend of source height with coronal luminosity is seen in which is inconsistent with a constant at $> 3\sigma$ (see Methods). The luminosity spans an order of magnitude, with the corresponding source height going from ~5 $R_g$ at the lowest fluxes to ~15 $R_g$ at the highest fluxes. The change in source height with $L_{2-10\ keV}$ is independent of spin (see Extended Data Figure 4).

The simultaneous model fitting provides strong constraints on key parameters; we find a high spin value, $a = 0.97^{+0.01}_{-0.13}$, a correspondingly small inner disc radius, $R_{in} = 1.7^{+1.1}_{-0.4} R_g$ and an inclination to the observer of, $i = 68^{+4}_{-6}°$ (see Extended Data Fig 5), consistent with the values obtained from broadband spectral fitting[19]. The posterior distribution of black hole mass is shown in Figure 3, with $M_{BH} = 1.9 \pm 0.2 \times 10^6 M_\odot$, in excellent agreement with $M_{BH} \sim 2 \times 10^6 M_\odot$ recently obtained[7] from the X-ray PSD break relation[26].

The geometry in our theoretical transfer function model is a simplified approximation of a real corona, which likely has some radial and vertical extent[12]. The effect of changing the corona's extent would be to change the low-frequency lag component[5] or add minor oscillatory structure to the reverberation component[6], however, the change in height of the corona dominates the changes to the lag spectrum. At high accretion rates, the disc could have some geometric thickness, changing the path length from corona to the disc, the effects of which are not expected to affect the current results[27]. Once these models are fully developed, X-ray reverberation will be a powerful method for determining the fine details of the accretion process.

Understanding how the disc-corona system couples together, how the corona is powered and their relation to jets, are longstanding problems in accretion physics[12,28]. Changes in coronal height coupled with long-term source flux have been observed using time-averaged spectra in similar AGN[29]. More recently, longer-term changes have been observed over a ~month timescale in a stellar-mass black hole X-ray binary, as it transitions through an accretion state[30]. Our analysis of IRAS 13224–3809 provides a further, unique insight into these innermost regions and their behaviour on far more rapid timescales (when measured in dynamical timescales, a day in IRAS 13224–3809 corresponds to less than a second for the smaller masses of black hole X-ray binaries). In addition to variations occurring within the corona on all timesclaes[7], we are also seeing changes in the coronal height occurring on less than the viscous timescale (days to months) of the inner accretion disc. The size of the corona inferred from our height estimates is consistent with the gravitational microlensing results[16].

In this work, we have been able to break the degeneracies in reverberation modelling by fitting multiple epochs during which the source height changes, providing a strong estimate of spin, inner disc radius inclination and black hole mass. The relative statistical error on $M_{BH}$ is comparable with that obtained for other AGN from the current state-of-the-art optical reverberation mapping method[9]. However, due to the weak optical variability in IRAS 13224–3809 and many other AGN[10], this traditional method cannot be applied in those cases, making the use of X-ray lag spectra a viable and powerful alternative. With reverberation lags detected in ~50 percent of nearby AGN[2,21], these results demonstrate how X-ray reverberation modelling will reveal important details about the inner ($\lesssim 50 R_g$) accretion flow and corona with future modelling capabilities and long-term monitoring campaigns. In IRAS 13224-3809, this reveals a dynamic picture of immediate region around the black hole, which is otherwise inaccessible.

**Corresponding author**
Correspondence and requests for materials should be directed to W. N. Alston (wna@ast.ca.ac.uk)

**Acknowledgements**
WNA and ACF acknowledge support from the European Research Council through Advanced Grant 340442, on Feedback. MLP and CP acknowledge support from ESA Research Fellowships. MD and MCG acknowledge support provided by the GA CR grant 18-00533S. MCG acknowledges funding from ESA through a partnership with IAA-CSIC (Spain). DJW and MJM appreciates support from an Ernest Rutherford STFC fellowship. DJKB acknowledges a Science and Technology Facilities Council studentship. CSR thanks the UK Science and Technology Facilities Council for support under Consolidated Grant ST/R000867/1. This research has been partially funded by the Spanish State Research Agency (AEI) Project No. ESP2017-87676-C5-1-R and No. MDM-2017-0737 Unidad de Excelencia "María de Maeztu" - Centro de Astrobiología (CSIC-INTA). GM also acknowledges funding by the Spanish State Research. Agency (AEI) Project No. ESP2017-86582-C4-1-R. BDM acknowledges support from the European Union's Horizon 2020 research and innovation programme under the Marie Skodowska-Curie grant agreement No. 798726. This paper is based on observations obtained with XMM-Newton, an ESA science mission with instruments and contributions directly funded by ESA Member States and the USA (NASA).

**Author contributions**.
WNA performed the data analysis and lag modelling, and wrote the manuscript. EK performed a complementary time lag data analysis. CP and JJ performed the time-averaged spectral modelling. MLP performed the rms-spectrum modelling. The remaining authors contributed to the discussion and interpretation.

**Competing Interests.**
The authors declare that they have no competing financial interests.

**Correspondence.**
Correspondence and requests for materials should be addressed to WNA (email: wna@ast.cam.ac.uk).

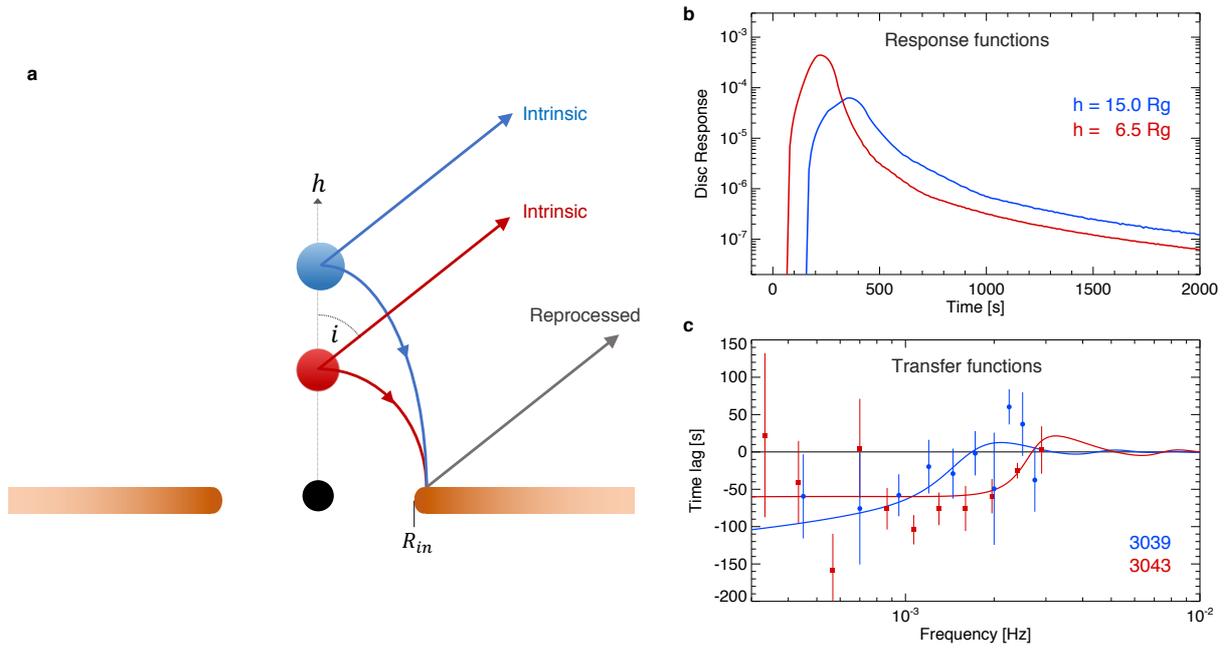

**Figure 1. The relativistic transfer function model used to fit the time lag spectra**. **a**. Schematic of the model geometry, seen at inclination $i$ by a distant observer. The X-ray corona (blue/red) at height $h$ irradiates the accretion disc which extends to some radius $R_{in}$ around the central black hole. The additional path length between the direct and reflected emission sets the reverberation timescale. The curved photon paths are caused by gravitational light bending. **b**. The disc response functions to a $\delta$-function flash of emission in the corona for two representative observations, 3039 and 3043, for the best-fitting model with black hole mass $M_{BH} = 1.9 \pm 0.2 \times 10^6\ M_\odot$, spin value $a = 0.97$ and heights $h = 15.0\ R_g$ (blue) and $h = 6.45 R_g$ (red). **c.** The best-fitting transfer function models to observations 3039 and 3043 (with $1\sigma$ error bars) after subtracting the additional intrinsic component (see Methods). The black line is the zero time-lag axis.

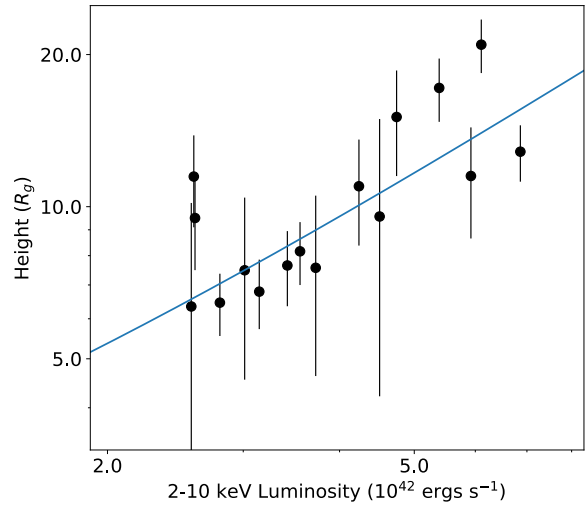

**Figure 2. The observed 2-10 keV luminosity of the intrinsic component versus the source height.** The data and $1\sigma$ error bars are fit using weighted linear regression (blue), which is inconsistent with a constant at $> 3\sigma$ level. The $1\sigma$ confidence region on the linear model is shown in grey.

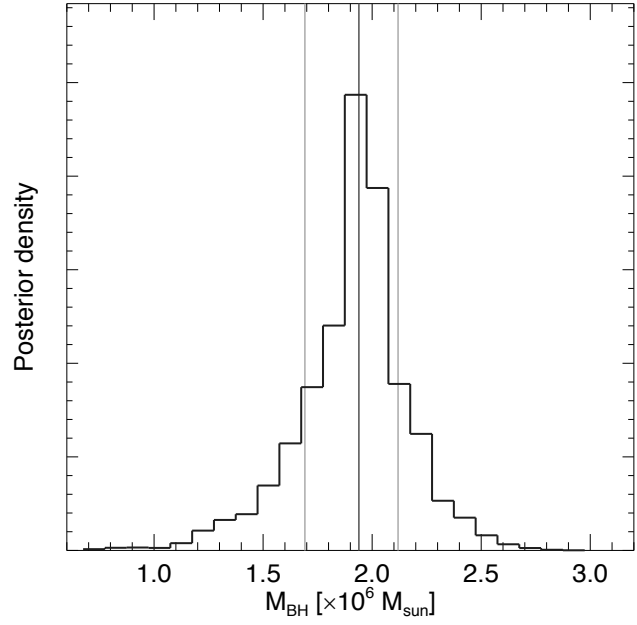

**Figure 3**. **The posterior distribution for black hole mass from the best-fitting model.** A value of $M_{BH} = 1.9 \pm 0.2 \times 10^6\ M_\odot$ is obtained. The median and 68% credible regions (equivalent to $1\sigma$ of a Gaussian distribution) are shown in grey.

# Methods

**Data reduction.**
We make use of the 16 full-orbit (~130 ks) *XMM-Newton* observations from 2011 and 2016, totalling 2 Ms observation time (Table 1). We use the EPIC-pn camera[31], due to its higher throughput. The Observation Data Files (ODFs) were processed following standard procedures using the XMM-Newton Science Analysis System (SAS; v15.0.0), using the conditions PATTERN 0-4 and FLAG = 0. The source counts were extracted from a circular region with a 20 arc seconds radius. The background was extracted from a large rectangular region on the same chip, placed away from the source and avoiding the Cu ring on the EPIC-pn chip. We produce background-subtracted light curves in the 0.3-1.0 keV (soft) and 1.2-5.0 keV (hard) bands (see[7] for more details). Details of the broadband spectral analysis can be found in[11,16]. The spectral modelling includes a component for the observed ultrafast outflow.

**Cross-spectral time lags.**
From two evenly sampled time series $x_t$, $y_t$ we can compute the complex-valued Fourier Transforms $X_f$, $Y_f$. We can estimate the power spectrum for each time series using $|X_f|^2$, $|Y_f|^2$, after applying some normalisation factor and subtracting the Poisson noise level. From these we compute the cross-spectrum $C_{xy} = X_f^* Y_f = |X_f||Y_f|e^{i(\phi_{y(f)} - \phi_{x(f)})}$, where $*$ denotes the complex conjugate. We estimated the cross-spectrum by first averaging the complex $C_{xy}(f)$ values over $m$ non-overlapping time series segments and then averaging in $n$ geometrically spaced frequency bins[2]. From the argument of $C_{xy}(f)$ the we obtain a phase lag $\phi(f) = \arg \langle C_{xy}(f) \rangle$ which may be transformed into the corresponding time lag $\tau(f) = \phi(f)/(2\pi f)$. This provides the (time averaged) frequency-dependent time lags between any correlated variations in $x_t$, $y_t$. Errors on $\tau(f)$ are estimated using standard formulae[32–35]. The segment lengths are typically 20 ks and are binned to have $m \times n > 20$ cross-spectral estimates per frequency bin, in order to have approximately Gaussian distributed errors.

From the cross-spectrum we can also obtain the coherence, a measure of the linear correlation between two time-series as a function of Fourier frequency[33]. The coherence is defined as $\gamma_{xy}^2(f) = |\langle C_{xy}(f) \rangle|^2 / [\langle |X(f)|^2 \rangle \langle |Y(f)|^2 \rangle]$, and takes on values in the range [0,1]. $\gamma_{xy}^2(f)$ is corrected for the effect of Poisson noise[33], providing an estimate of the intrinsic coherence. The intrinsic coherence is a biased estimator above some frequency $f_{max}$, above which coherence and time-lag measurements become unreliable[35]. We therefore limit the analysis to frequencies below $f_{max} = 2.8 \times 10^{-3}$ Hz.

**Model fitting.**
The energy bands used to make the light curves contain a mixture of both the intrinsic (power-law) and reprocessed (reflection) components (because they have a 'continuum' over a broad energy range – the so called 'dilution' effect). With current instruments, we are unable to produce fast cadence light curves for the intrinsic and reprocessed component, as is the case for the optical reverberation technique[29]. X-ray reverberation modelling is therefore performed in Fourier space (e.g. see[2]), starting with a model setup and deconvolving it through the data to make inferences.

We jointly fit all 16 lag-frequency spectra simultaneously within XSPEC using the KYNREVERB model[22,36]. This uses a fully relativistic ray-tracing code in a vacuum to calculate photon paths from the corona to the disc and to the observer, as well as the reflected photons from the disc to the observer. The re-processing in an ionised accretion disc is self consistently calculated at each radius from REFLIONX tables[37]. The observed intrinsic emission and reflection component are used to determine the dilution in each energy band[2]. We fix some parameters in the model for individual spectra based on the values obtained from the time averaged spectral fits to each observation[11,16]. The Fe abundance is fixed at Z = 3 and the disc density at $n_d = 5 \times 10^{17}$ cm$^{-3}$. The observed 2-10 keV luminosity, $L_{2-10keV}$, and the primary continuum photon index, $\Gamma$, are listed in Table 1. The intrinsic lag component[38–41] is modelled as $\tau = kf^{-q}$, which is included as an additional component in the modelling. $M_{BH}$, spin $a$ and inclination $i$, are free parameters and tied across the lag spectra.

Source height $h$ and the intrinsic component parameters $k$ and $q$ are untied and free to vary. We tie $R_{in}$ across the observations, which is a free parameter and free to vary down to, but not fixed at $R_{ISCO}$. The outer disc radius is fixed at 400 $R_g$, and the model fit is insensitive to this parameter.

The model fitting to all 16 lag-frequency spectra was explored using a Bayesian framework, treating the log-likelihood function as $\sim(\chi^2/2)$. Uniform and wide priors were assigned to the parameters. The posterior distribution of the parameters was determined using Markov Chain Monte Carlo (MCMC) to draw sets of parameters at random. We use the Goodman & Weare affine invariant MCMC ensemble sampler, within the python module EMCEE[42], which is implemented in XSPEC through XSPEC EMCEE. We generated 200 chains of length 50,000 after an equal length burn-in period. Convergence is assessed using the Gelman-Rubin test[43]. Posterior parameter values are derived from the median and 68% credible regions (equivalent to 1-sigma of a Gaussian distribution).

The best fitting model gives $\chi^2 = 224$ for 126 degrees-of-freedom (d.o.f). The MCMC parameter draws for each posterior are shown in the scatter plots in Extended Data Figure 5. These plots show how there is no degeneracy in the parameters $M_{BH}$, spin, $R_{in}$, $i$ and source height $h$ when the model is simultaneously fit over multiple lag spectra. In Extended Data Figure 6 we show the $M_{BH}$ and $h$ posterior distributions when the model is fit to individual lag spectra. We show a representative low flux observation (3049) and high flux (3050), where the degeneracy between parameters can be seen, which is typical of fitting single lag spectra only.

The source height $h$ and observed 2-10 keV luminosity values were fit with a linear model $h = \beta_0 + \beta_1 L^{42}_{2-10keV}$, where $\beta_0$ and $\beta_1$ are coefficients and $L^{42}_{2-10keV}$ is the observed 2-10 keV luminosity in units of $10^{42}$ ergs$^{-1}$. We use simple weighted least squares regression, with the weighting equal to the reciprocal of the variance of the measurements. For the free $R_{in}$, $i$ and spin model shown in Figure 2, we find $\beta_0 = 1.2 \pm 1.8$ and $\beta_1 = 2.1 \pm 0.46$. The slope parameter of the is inconsistent with a constant, with $p = 0.0005$ from a Student $t$-test.

We test whether the height-luminosity relation is independent of spin using 3 fixed values of $a = 0.998$, $a = 0.7$ and $a = 0$, and fixing the inner disc edge to the corresponding $R_{ISCO}$. By assuming that the disc extends down to $R_{ISCO}$, the spin parameter therefore determines the distance between the inner edge and the source height. For $a = 0.998$ the fit gave $\chi^2 = 265$ for 134 d.o.f. The case with $a = 0.7$ gives $\chi^2 = 270$ for 134 d.o.f, and the case with $a = 0$ gives $\chi^2 = 277$ for 134 d.o.f.

For all three cases we find the linear model fit to the respective $h$ versus $L^{42}_{2-10keV}$ is inconsistent with a constant at > 3$\sigma$. For the $a = 0.998$ case, fitting gave parameter estimates $\beta_0 = 1.6 \pm 1.5$ and $\beta_1 = 1.89 \pm 0.43$. The slope parameter is inconsistent with a constant, with $p = 0.0006$ from a Student $t$-test. For the $a = 0.7$ case, we derive $\beta_0 = 2.3 \pm 1.5$ and $\beta_1 = 1.33 \pm 0.34$, with $p = 0.0017$.
$\beta_0 = -0.1 \pm 1.8$ and $\beta_1 = 2.26 \pm 0.59$, with $p = 0.0019$.

For each model case for spin (free, fixed at 0, 0.7 and 0.998) we test the robustness of the $h$ measurements by repeating the lag model fitting procedure, allowing $\Gamma$ in the reflection component to vary. The average error is $\Delta\Gamma \sim 0.15$. We check the height-luminosity relation is independent of inclination, by testing each spin case with inclination fixed a $i = 30°$ and $i = 45°$. We test for correlations of the intrinsic component parameters, $k$ and $q$, with $L^{42}_{2-10}$, but find these are consistent with zero correlation. We inspect the MCMC output to ensure the intrinsic component is not degenerate with any of the reverberation parameters.

Although the position of the inner edge of the accretion disc is not expected to change on day timescales, to test whether changes in $R_{in}$ could produce a similar result to a changing $h$, we repeat the lag-spectral fitting with $R_{in}$ untied across all 16 observations. A small inner radius is favoured in all spectra and the dependence on $h$ with intrinsic luminosity is still observed. Tying $h$ across all lag spectra and allowing $R_{in}$ to vary results in a significantly worse fit, with $\chi^2 = 369$ for 126 d.o.f, giving $\Delta\chi = 145$ for the same d.o.f compared to the preferred model.

**Modelling the rms-spectrum.**
We investigated the energy dependence of the variability using the rms-spectrum[44,45]. The rms in a given energy band was calculated in absolute units by integrating the Poisson noise subtracted power spectra (PSD) over the frequency range of interest. This is equivalent to calculating the rms in the time-domain, where the lower frequency-bound is equivalent to 1/(*segment length*), and the upper frequency-bound is equivalent to 1/2*dt*, where *dt* is the time bin used[44]. The rms-spectra were then calculated in fractional units by dividing by the time-averaged spectrum.

We calculated the rms-spectrum in three frequency bands: [0.08,6.0] x10$^{-4}$ Hz (broad); [1.0,5.0] x10$^{-5}$ Hz (low-frequency, LF) and [4.0,20.0] x10$^{-4}$ Hz (high frequency, HF). These are shown in Extended Data Figure 7, along with the IRAS 13224-3809 PSD indicating the frequency ranges used. The HF band picks out the reverberation timescale (see also Extended Data Figure 1), whereas the LF is dominated by the intrinsic variability. Due to the PSD shape, the broad band is dominated by the low frequency component.

To simulate the rms-spectrum we used a simplified version of the model from[17], consisting of a power-law, reflection, and a black-body, all absorbed by an ionised outflow (these variable components are also seen using principle component analysis[46,47]). We relate the parameters according to the correlations identified by[17]: the reflection flux, $F_{ref}$, is proportional to the power-law flux by $F_{ref} \propto F_{pow}^{0.5}$, the ionization of the absorber, $\xi_{abs}$, is related by $\xi_{abs} \propto F_{pow}$, and the power-law index scales as $\Gamma \propto 0.6 \log(F_{pow})$. We vary the power-law flux from a log-normal distribution, with width, $\sigma = 0.5$ dex. We also vary the temperature of the black-body, following a linear normal distribution, with $\sigma = 0.03$ keV. We fix the flux of the black body to follow the observed $T^4$ disc temperature relation[17]. We generated 1000 spectra based on this model, and calculated the fractional variance. The simulated models are shown in Extended Data Figure 7, which describe the data well.

The broad and low-frequency rms-spectra are described by a highly-variable absorbed power-law, with slower varying reflection features and an even slower blackbody. This is the origin of the drop is variability power below ~1 keV and above ~1.3 keV. Apparent in the broad and low-frequency rms-spectra data is a variable feature at 8 keV, which is caused by the variable ultrafast-outflow absorption feature[10,11]. Because this structure in the Fe band is mainly due to absorption, the Fe abundance and relativistic blurring parameters have negligible impact on the shape of the rms-spectrum in this band (i.e. super-solar Fe abundance is not required[48]).

For the HF rms-spectrum an increase in reflection variability is required, with $F_{ref} \propto F_{pow}^{0.8}$, in order to model the flatter energy dependence above 1 keV. This indicates that the reflection component and power-law are varying more coherently, as is expected from the disc reflection scenario.

To highlight the effect of the variable ultrafast outflow in each band, we repeat the simulations without this component (dashed lines in Extended Data Figure 7). It can be seen that the UFO contributes ~1-2% to the total variability amplitude, with this fraction decreasing at higher frequency bands.

**Alternative model for AGN variability.**
We note that there is an alternative model in the literature to attempt to explain the short-timescale (≲ 10 ks) variability in AGN[48-51]. The partial covering model proposes that the observed fast X-ray variability is produced by a series of intervening eclipsing clouds between the corona and the line of sight to the observer, located at >100s $R_g$ from the central source[49]. These models can also include an additional ionised outflow located at 50-500 $R_g$[ref 50], which produces Fe K emission by scattering off ions in the outflow. In each case, the central X-ray source is located within ~20 $R_g$, which is also consistent with the coronal extent in the disc reflection scenario.

Whilst these models are able to reproduce some aspects of the spectral variability, they fail to describe the fundamental properties of the variability process observed in accreting systems: the lognormal distribution of fluxes and the linear rms-flux relation[7,13,52-54]. The rms-flux relation is a universal signature of an accretion disc and is observed in many wavebands and in all types of accreting compact objects, such as non-jetted AGN, but also including accreting white dwarfs, young stellar objects, as well as neutron star and black hole X-ray binaries[7,13,55-63]. Magneto-hydrodynamic (MHD) accretion disc simulations are able to reproduce the rms-flux relation[64,65].

One important factor of the rms-flux relation is that it is observed over a broad range of timescales[53]. This tells us that the different variability timescales are coupled together, meaning the underlying variability process is multiplicative rather than additive[53]. The leading model to describe this is the propagation of accretion disc fluctuations, where slower variations at larger disc radii propagate inward and modulate the faster variations produced at smaller radii[53,66-70]. This observation rules out additive models for the origin of the fast variability. For the rms-flux relation to be produced by the partial covering scenario requires a highly contrived situation where the clouds of different sizes are perfectly aligned and rotating at different radii, and for this to be the case in all AGN, which are known to show a variety of absorption signatures from obscuration of the central corona[8,71]. This linear relation is observed on both long (>10 ks) and short (up to 100 s) timescales in IRAS 13224-3809[7,13].

**Data availability statement**. The data that support the plots in Figure 2 and Extended Data Figures 1 and 4 and included as source data in the Supplementary Information. All other data used in figures within this paper and other findings of this study are available from the corresponding author upon request. All data used in this work is publicly available. The *XMM-Newton* observations can be accessed from the *XMM-Newton* science archive (http://nxsa.esac.esa.int/nxsa-web/).

**Code availability.** All the code used for the data reduction is available from their respective websites. XSPEC is freely available online. The transfer function model KYNREVERB is available at https://projects.asu.cas.cz/stronggravity/kynreverb. The MCMC sampler EMCEE is available at http://emcee.readthedocs.io/en/stable/index.html, with the XSPEC implementation available at https://github.com/jeremysanders/xspec_emcee.

## References cont.

# Extended Data

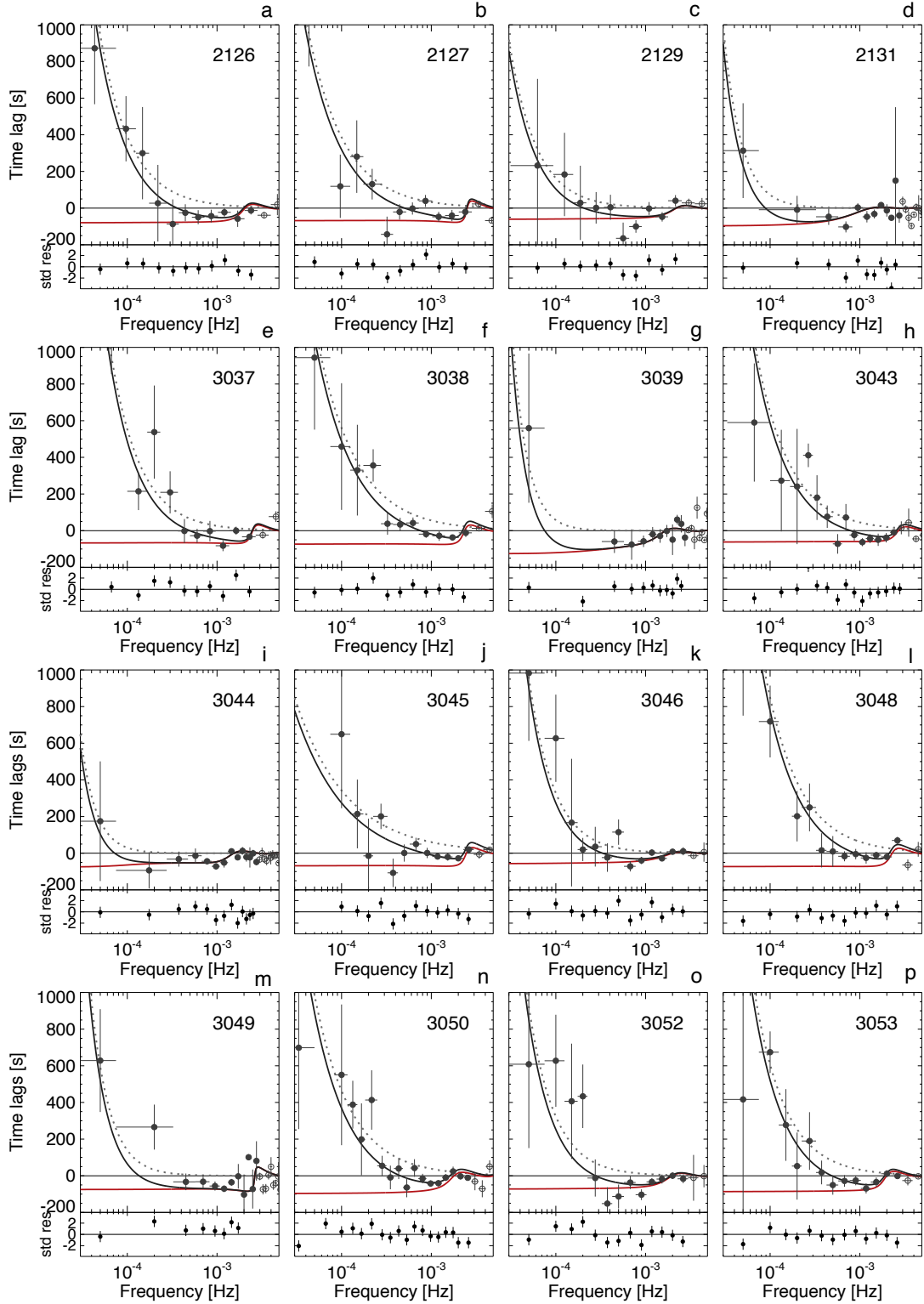

**Extended Data Figure 1. Lag frequency spectra for the 16 *XMM-Newton* orbits.** The best-fitting transfer function model, where $M_{BH}$, spin $a$, $R_{in}$ and inclination $i$ are tied, but free, is shown in black solid line. The reverberation component is shown in red and the intrinsic component is shown in dotted grey. The zero time-lag as a function of frequency is shown as a black horizontal line. The black circles are the data with their $1\sigma$ error bars. The time-lag estimate at $2\times10^{-4}$ Hz in observation 3039 has $\tau = 307 \pm 87$ s, but is clipped from the plotting region. The lower panels show the standard residuals.

| XMM-Newton orbit *no.* | $L_{2-10keV}$ ($10^{42}$ ergs$^{-1}$) | $\Gamma$ | Height ($R_g$) |
|---|---|---|---|
| 2126 | 4.50 | 2.45 | $9.6^{+5.3}_{-3.3}$ |
| 2127 | 3.72 | 2.45 | $7.6^{+2.9}_{-2.3}$ |
| 2129 | 2.59 | 2.40 | $9.5^{+2.0}_{-1.3}$ |
| 2131 | 6.10 | 2.40 | $20.9^{+2.5}_{-3.4}$ |
| 3037 | 3.01 | 2.28 | $7.5^{+2.9}_{-2.6}$ |
| 3038 | 3.42 | 2.52 | $7.6^{+1.3}_{-1.8}$ |
| 3039 | 4.74 | 2.56 | $15.0^{+3.6}_{-3.4}$ |
| 3043 | 2.79 | 2.11 | $6.5^{+0.9}_{-0.9}$ |
| 3044 | 5.38 | 2.47 | $17.2^{+2.5}_{-3.0}$ |
| 3045 | 3.14 | 2.40 | $6.8^{+1.1}_{-1.6}$ |
| 3046 | 2.58 | 2.26 | $11.5^{+2.4}_{-1.3}$ |
| 3048 | 3.55 | 2.52 | $8.2^{+1.2}_{-2.2}$ |
| 3049 | 2.56 | 2.00 | $6.3^{+3.8}_{-1.5}$ |
| 3050 | 6.86 | 2.68 | $12.8^{+1.6}_{-2.4}$ |
| 3052 | 4.23 | 2.48 | $11.0^{+2.6}_{-2.2}$ |
| 3053 | 5.92 | 2.65 | $11.5^{+2.9}_{-1.6}$ |

**Extended Data Fig 2. Table of source properties and best-fitting model posteriors**. Column 1 shows the XMM-Newton orbit number. Column 2 shows the 2-10 keV luminosity in units of $10^{42}$ ergs$^{-1}$. Column 3 shows the photon index $\Gamma$ from the time-averaged spectral model fits. Column 4 shows the posterior values for source height h from best-fitting the transfer function model fit with parameters $R_{in}$, inclination *i* and spin free and tied across the lag spectra.

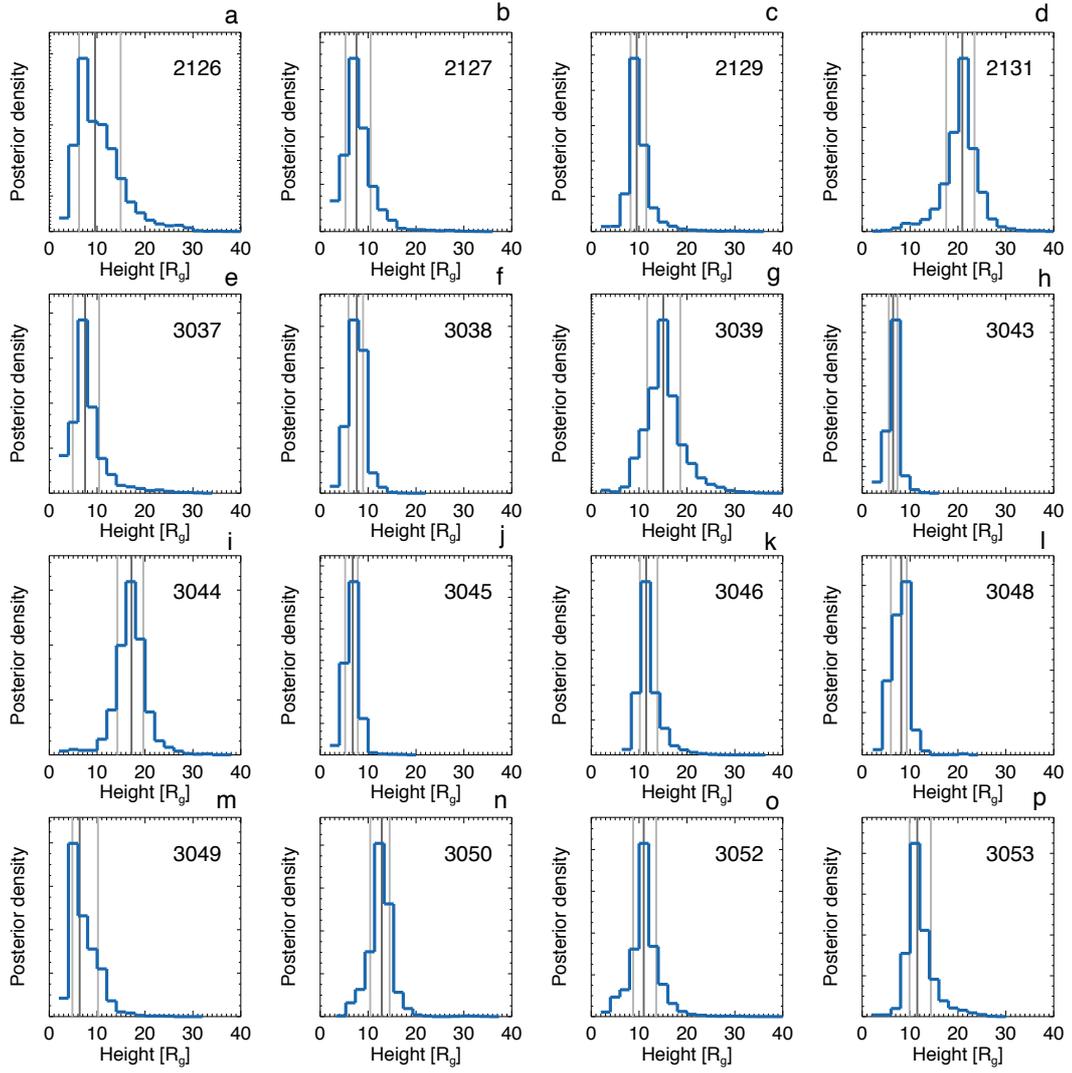

**Extended Data Figure 3. Posterior distributions for source height from the best fitting model**. The model has parameters $M_{BH}$, $R_{in}$, inclination $i$ and spin $a$, free and tied. The median and 68% credible regions (equivalent to $1\sigma$ of a Gaussian distribution) are shown in grey.

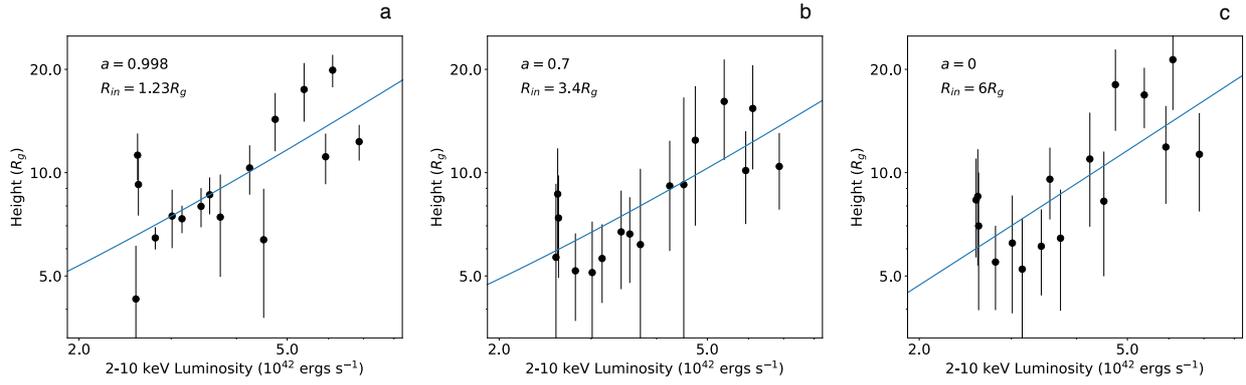

**Extended Data Figure 4**. $L_{2-10\ keV}$ **luminosity vs source height $h$ from the model fit with fixed spin.** Panel **a** shows the spin value $a = 0.998$, panel **b** shows $a = 0.7$, and panel **c** shows $a = 0$. The corresponding fixed inner disc radius $R_{in}$ is stated. The solid line is the best-fitting linear regression model together with the $1\sigma$ confidence region on the model. The error bars on individual data points are $1\sigma$ (see Methods).

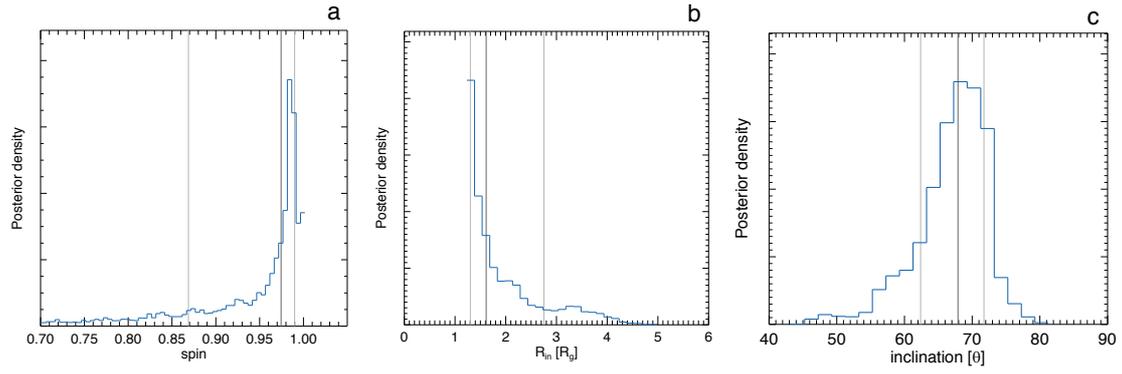

**Extended Data Figure 5**. **MCMC posterior densities for the model fit with parameters free**. Panel **a** shows the spin *a*, panel **b** shows the inner disc radius $R_{in}$, and panel **c** shows the inclination *i*. The median and 68% credible regions (equivalent to 1-sigma of a Gaussian distribution) are shown in grey vertical lines.

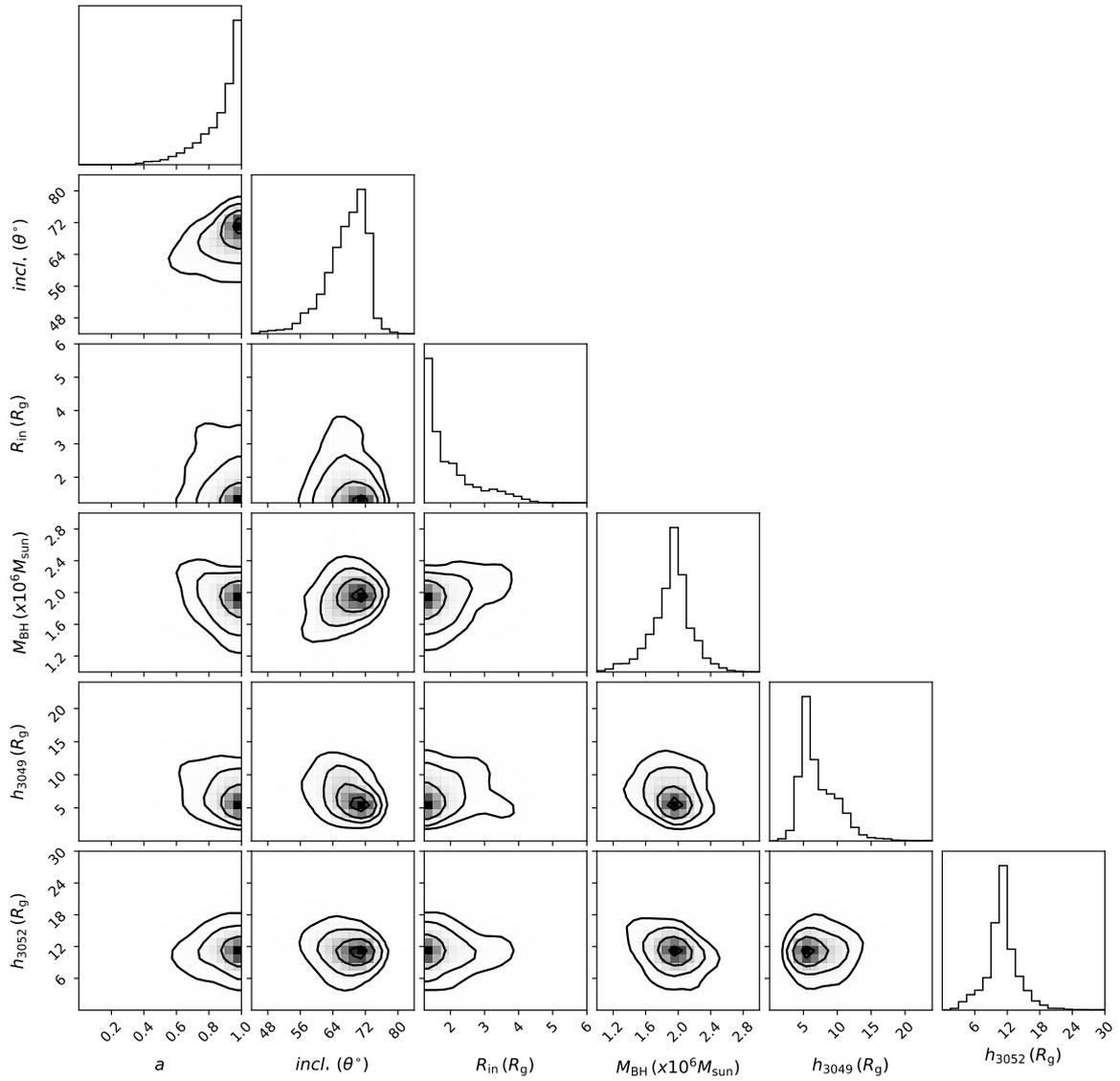

**Extended Data Figure 6. Scatter plots for the MCMC posterior parameter distributions.** Shown are the posteriors for spin $a$, inclination $i$, $R_{in}$, $M_{BH}$, as well as source height $h$ for two representative observations, with a low (3049) and high (3052) source flux.

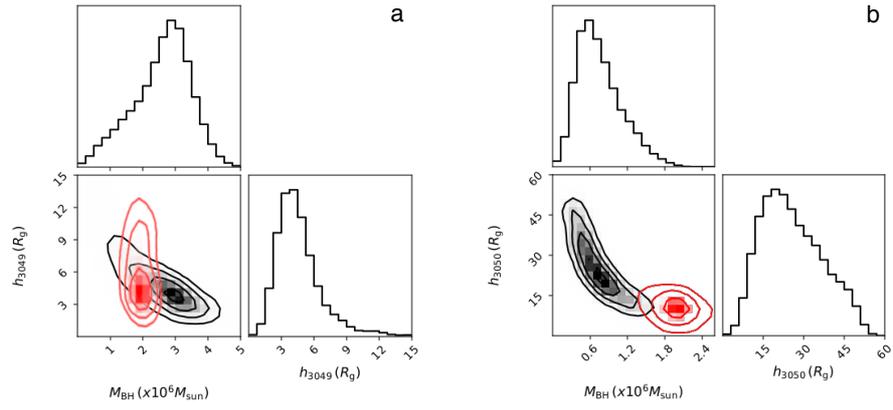

**Extended Data Figure 7**. **The posterior parameters $M_{BH}$ and source height $h$ from the model fit to just one individual lag spectra.** Panel **a** shows the model fit to observation 3049 and panel **b** that for observation 3050. The degeneracy between the model parameters can be seen. The red contours are the MCMC posteriors for the joint fit.

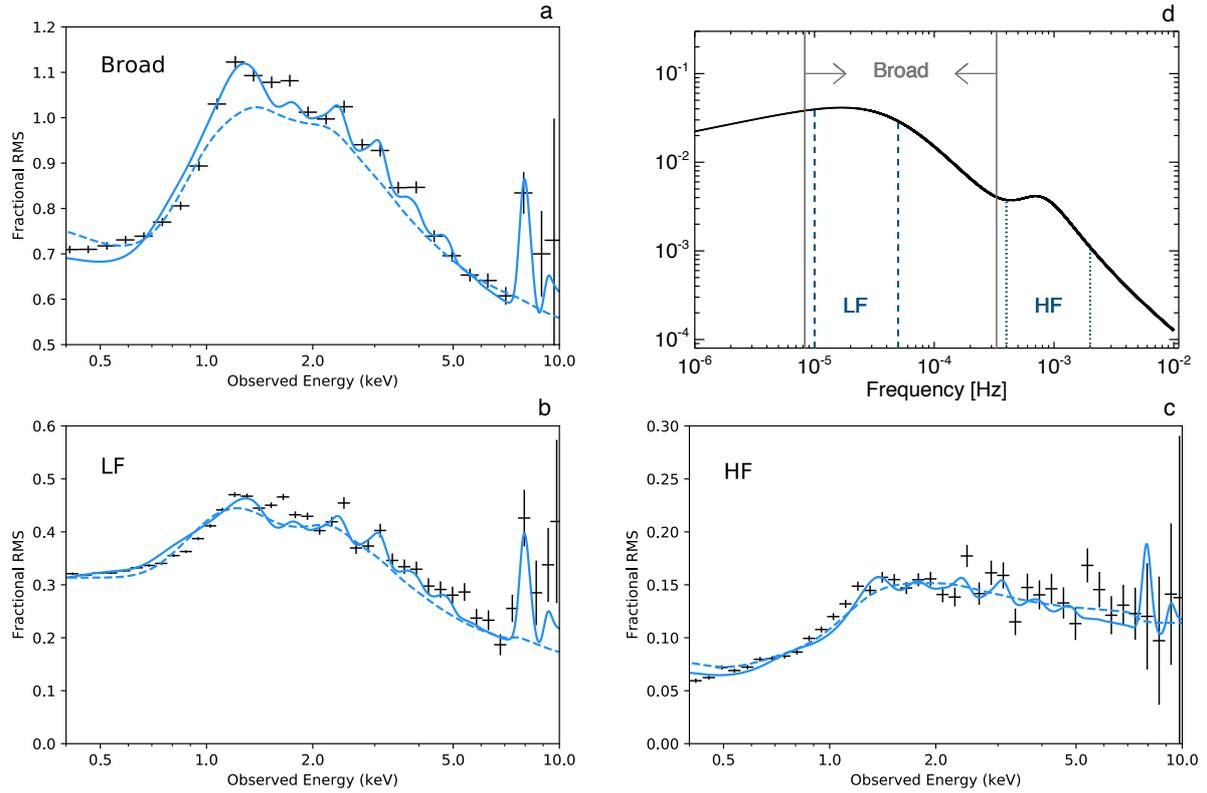

**Extended Data Figure 8. Modelling the rms-spectrum with the disc reflection scenario.** The solid blue lines are the model with the inclusion of the ultrafast outflow (UFO) and the dashed blue lines are without the UFO (see Methods for details). Panel **a** shows the 'Broad' rms-spectrum calculated from $[0.08, 6.0] \times 10^{-4}$ Hz. Panel **b** shows the rms-spectra for a low frequency band, LF = $[1.0, 5.0] \times 10^{-5}$ Hz. Panel **c** shows that of the high frequency band, HF = $[4.0, 20.0] \times 10^{-4}$ Hz. Panel **d** shows the power spectral density (PSD) for the data, with the frequency ranges used for the rms-spectra indicated by the vertical solid (Broad), dashed (LF), and dotted (HF) lines. The HF band is where the reverberation signal dominates. $1\sigma$ error bars are shown on all data points.